\documentclass{article}
\usepackage{spconf,amsmath,amssymb,graphicx,booktabs,hyperref}
\hypersetup{hidelinks}
\usepackage{microtype,balance}
\title{Temporal Anchors and Editing Sensitivity in Partial Speech Spoofing: A Controlled Study}
\name{Xiaosu Su, Yun Cao, Yiping Ni, and Xiaowei Yi}
\address{Institute of Information Engineering, Chinese Academy of Sciences, Beijing 100085, China\\
School of Cyber Security, University of Chinese Academy of Sciences, Beijing 100085, China}
\begin{document}
\maketitle
\begin{abstract}
Partial-spoof detectors must reject synthetic content while accepting benign edits. We diagnose temporal anchors and boundary-consistency training using frozen WavLM features, nonlinguistic unit controls, and source-locked evaluation. Genuine--genuine and genuine--fake splices contrast editing false alarms with synthetic-content misses; they do not isolate a unique causal artifact. With layer-6 features, phone units have higher PartialSpoof evaluation frame EER than soft frames (11.18\% versus 9.59\%). In 500 retrospective PS-eval cases, augmentation reduces genuine-splice false alarms; synthetic-core misses increase at 5\% PS-development FPR but not demonstrably at 1\%. A 200-case Llama A follow-up retains the false-alarm reduction but does not establish a miss-rate increase. Constructed-case ranking can improve while fixed-threshold misses rise. Consistency gives no uniform gain, and encoder-layer rankings change across corpora. We report frame and event localization with three seeds, including word units, and treat PS evaluation as retrospective. Audit code, selected training scripts, and result summaries are available at \url{https://github.com/mysxs/partial-spoof-diagnostics}.
\end{abstract}

\section{Introduction}
A small synthetic insertion can change an utterance without replacing its surrounding genuine speech. PartialSpoof~\cite{partialspoof}, Half-Truth Audio Detection (HAD)~\cite{had}, and LlamaPartialSpoof~\cite{llama} expose this problem under different generation and editing conditions. A useful detector must localize manipulated content while accepting genuine edits. A splice discontinuity is therefore an ambiguous cue: it can occur with either genuine or synthetic material.

Linguistic intervals offer an attractive prior. Pooling within phones or words could suppress incidental frame fluctuations, and predictions outside edited boundaries could be regularized to remain invariant under smoothing. However, neither intuition guarantees better localization. A short fake interval may fall within a phone; a word may combine genuine and fake content; and smoothing can change acoustic evidence beyond a single discontinuity. Improvements relative to an unaugmented model also do not isolate consistency from exposure to additional examples.

We ask three linked questions. First, do temporal anchors improve detection without sacrificing short-event localization? Second, after holding the augmentation family fixed, does a consistency loss change the balance between benign-edit false alarms and synthetic-content misses? Third, do source-selected scores transfer across corpora and operating points? We contribute (i) matched frame, phone, word, and nonlinguistic controls; (ii) paired GG/GF interventions that expose the two error types separately; and (iii) source-locked cross-corpus and threshold-transfer analysis. The paper is diagnostic: it does not introduce a new encoder or claim that the tested anchor or consistency designs are universally ineffective.

\section{Related Work}
PartialSpoof provides segment annotations and multi-resolution countermeasures~\cite{partialspoof}. HAD studies partially fake recordings in Chinese~\cite{had}; LlamaPartialSpoof extends evaluation to generated disinformation and multiple editing procedures~\cite{llama}. These corpora make localization and transfer distinct requirements.

CFPRF combines frame detection and proposal refinement~\cite{cfprf}. Its reproduced-model analysis shows that in-domain EER and out-of-domain utility may disagree~\cite{robust}. We place this study alongside ASVspoof/WaveFake benchmarks~\cite{asvspoof19,asvspoof21,wavefake} and earlier ASVspoof protocols~\cite{asvspoof15,asvspoof17}. Prior spoofing methods include CQCC features~\cite{cqcc,cqccselect} and raw/graph architectures~\cite{rawnet2,gat,aasist}; RawNet originated as a speaker-verification encoder~\cite{rawnet}. WavLM, wav2vec~2.0, HuBERT, and XLS-R represent the self-supervised family~\cite{wavlm,wav2vec2,hubert,xlsr}; audio studies document front-end sensitivity and cross-domain shifts~\cite{sslfront,generalization,hardergeneralization,crossdomain,domaingen,continual}. Our paired GG/GF interventions contrast benign-edit false alarms with missed synthetic content, while augmentation-only controls separate exposure from consistency; challenge, adversarial, and multilingual studies motivate transfer evaluation~\cite{add2022,advpost,advvc,vcartifacts,mlaad,tcn}.

\section{Controlled Formulation}
\subsection{Frame and linguistic predictions}
Let $x$ be a waveform and $y(t)\in\{0,1\}$ indicate synthetic content. A frozen WavLM Large encoder produces features $h_k$ at 20-ms stride. Layer normalization and a projection to 128 dimensions precede three residual temporal convolutions with kernel size 5 and dilations 1, 2, and 4. A linear output and sigmoid give spoof probabilities. The single-stream temporal head has 379,521 trainable parameters.

We distinguish official hard frame targets from soft occupancy targets. For time cell $C_k$, occupancy is
\begin{equation}
q_k=\frac{1}{|C_k|}\int_{C_k}y(t)\,dt.
\end{equation}
Soft frame targets average the two constituent official 10-ms labels after conversion to spoof-positive polarity. Development evaluation retains official hard 20-ms labels. Logits are interpolated using physical centers (first center 12.5 ms, stride 20 ms), with constant edge extension, rather than resizing sequences by length.

For aligned unit $U_j=[a_j,b_j)$, the unit model pools acoustic features and predicts occupancy $p_j$. Its target is $q_j=|U_j|^{-1}\int_{U_j}y(t)\,dt$. Training weights binary cross-entropy by duration:
\begin{equation}
\mathcal L_{\mathrm{unit}}=\frac{\sum_j |U_j|\,\mathrm{BCE}(p_j,q_j)}{\sum_j |U_j|}.
\end{equation}
Unit probabilities are projected to evaluation cells by overlap integration:
\begin{equation}
s_k=\frac{1}{|C_k|}\sum_j |C_k\cap U_j|p_j.
\end{equation}
A unit may thus be partly fake. Phone and word boundaries come from Whisper transcripts~\cite{whisper} followed by Montreal Forced Aligner~\cite{mfa}. Uncovered gaps are partitioned into intervals no longer than 80 ms. Missing alignments use an 80-ms grid, preserving all trials; PS evaluation includes 236 fallback utterances.

We also concatenate phone-pooled context to frame features and compare it with fixed-80-ms context and a duplicated-frame control. Separately, unit-output controls use fixed 80-ms intervals or deterministically permute each utterance's phone widths. The permutation preserves the unit-length multiset but moves boundaries independently of labels. These controls use the same frozen final-layer features, unit head, duration-weighted occupancy loss, and source selection as the phone-unit arm. Context models are distinct from predicting one score per unit.

\subsection{Matched boundary interventions}
We cross two representations (frame and phone) with three training conditions: original data, augmented data, and the same augmented data plus consistency. Genuine hosts receive genuine donors (GG) or synthetic donors (GF). Raw splices have smoothed variants. Training uses 5- and 20-ms smoothing; evaluation additionally uses 50 ms as a strength diagnostic. All training hosts and donors come from the training split.

The consistency condition minimizes
\begin{equation}
\mathcal L=\mathcal L_o+\lambda_a\mathcal L_a+
\frac{\lambda_c}{|M|}\sum_{k\in M}
\mathrm{JS}\!\left(\mathrm{Ber}(s_k),\mathrm{Ber}(s'_k)\right),
\end{equation}
where $\mathcal L_o$ and $\mathcal L_a$ are supervised losses on original and augmented data. Mask $M$ excludes a fixed 60-ms neighborhood of boundaries. Mixture occupancy supervises augmented boundaries; the core compares unchanged targets. We use $\lambda_a=1$ and $\lambda_c\in\{0,1\}$. Augmentation-only and consistency models receive the same intervention family. Smoothing is not assumed to remove all editing artifacts; the core mask does not guarantee that encoder features are unaffected.

\subsection{Localization diagnostics}
Predicted events are consecutive positive 20-ms cells at the saved development threshold. One-to-one matching maximizes eligible matches at intersection-over-union (IoU) $\geq0.5$, then total IoU. Missed and false events remain in recall and precision. An auxiliary offset head permits endpoint shifts of at most 200 ms. Classification scores and corrected proposals are evaluated separately: a lower classification EER does not establish better boundaries.

\begin{table*}[t]
\centering
\caption{20-ms frame EER (\%). Probe rows give three-seed mean $\pm$ sample SD. Layer 6 is selected on PS development data; layer 12 is a reported comparison, not a target-selected replacement. Released reCFPRF has a different encoder/training recipe and is one checkpoint.}
\label{tab:transfer}
\setlength{\tabcolsep}{6pt}
\begin{tabular}{lccccc}
\toprule
System & PS dev & PS eval & HAD test & Llama A & Llama B \\
\midrule
WavLM final, hard frame & $5.330\pm0.127$ & $11.123\pm0.020$ & $38.589\pm1.009$ & $42.413\pm0.444$ & $38.587\pm0.119$ \\
WavLM layer 6, hard frame & $1.738\pm0.043$ & $9.772\pm0.487$ & $56.582\pm5.003$ & $41.493\pm0.638$ & $41.245\pm0.069$ \\
WavLM layer 12, hard frame & $2.716\pm0.094$ & $8.278\pm0.418$ & $38.735\pm3.323$ & $40.815\pm0.177$ & $35.701\pm0.089$ \\
\midrule
Released reCFPRF & --- & 9.826 & 14.862 & 41.673 & 40.512 \\
\bottomrule
\end{tabular}
\end{table*}

\section{Experimental Protocol}
PartialSpoof (PS) supplies 25,380 training, 24,844 development, and 71,237 verified evaluation waveforms. Two stale evaluation-list entries absent from the verified archive are recorded separately; labels remain unchanged. External testing uses all 9,072 HAD test utterances and both LlamaPartialSpoof archives: A has 76,228 utterances and B has 64,388. A includes genuine, fully fake, and crossfaded partially fake speech; B contains partially fake speech constructed by cut/paste or overlap/add. A and B are archive designations, not new training splits.

Probes use seeds 1--3, eight epochs, batch size 32, and AdamW with learning rate $3\times10^{-4}$ and weight decay $10^{-4}$. Checkpoints and operating thresholds are selected on PS development data. External results do not select epochs, recalibrate thresholds, or reverse score polarity. The temporal-anchor comparison uses layer 6, selected on PS development for the hard-frame arm. Unit-output width controls and the 18 matched intervention models use final-layer features; their absolute rates therefore should not be compared directly with the layer-6 anchor table. We also report layer 12 as a representation comparison. Variability is sample standard deviation across seeds, not a confidence interval. Released reCFPRF is one checkpoint and a contextual reference, evaluated without retraining or its proposal-refinement stage.

The primary scalar is pooled \emph{20-ms frame EER} from the official kernel; fixed-threshold FPR/FNR use the PS-development threshold. HAD test and Llama B have no wholly genuine utterances, so utterance EER is undefined there. Initial controls contain 500 development-derived cases from 10 speakers; supplementary controls contain 500 new cases from 48 eval speakers (4,500 waveforms) disjoint from train/dev. The 18 final-layer controls, construction rules, 400-ms same-speaker donors, 60-ms core mask, Whisper/MFA alignments, and 0.1--20\% development-FPR grid were frozen before scoring. We report original dev-EER and dev-calibrated 1\%/5\% operating points; 2,000-draw speaker-and-seed intervals are pointwise and unadjusted. Because prior PS eval results informed the study, PS eval is retrospective validation of frozen predictions, not a prospectively unseen benchmark; no eval labels, thresholds, or checkpoints changed training or selection.
\textbf{Reproducibility.} We retain run configs and hashes for lists, labels, checkpoints, predictions, metric code, and event matching. Python/NumPy/PyTorch seeds and deterministic cuDNN are fixed; WavLM extraction uses 16-kHz mono audio, 20-ms centers (12.5-ms first), float32 heads, and no test-time augmentation. The release package contains these entry points without raw data, weights, credentials, or private paths.

\section{Results and Analysis}
\subsection{Temporal anchors do not resolve the localization trade-off}
Table~\ref{tab:anchors} compares layer-6 checkpoints: phone units increase EER and reduce event F1 and short-event recall relative to soft frames. Context arms establish no consistent gain. With final-layer unit outputs and the same loss, PS-eval EER is $11.237\pm0.224\%$ for phones, $11.977\pm0.284\%$ for permuted phone widths, and $11.189\pm0.127\%$ for fixed 80-ms units. Phones beat width permutation but not the fixed grid; this does not isolate linguistic information.

On full PS development, final-layer soft-frame, phone-, and word-unit models recover 51.15\%, 12.81\%, and 9.05\% of events shorter than 100 ms at IoU 0.5, retaining non-speech events and reference tails. Label-informed endpoint shifts bounded by 200 ms recover 74.133\% for phones versus 75.372\% for an 80-ms grid; this is a structural diagnostic, not a learned detector or a relaxed-IoU upper bound.

Table~\ref{tab:localization} audits final-layer localization with saved PS-development thresholds on retrospective PS eval. Word-unit seed-level EER/F1/recall is 12.696/84.420/82.141\%, 12.331/84.981/83.307\%, and 12.516/84.469/82.523\%.

\begin{table}[t]
\centering
\scriptsize
\caption{Complete PS-eval localization audit from frozen final-layer predictions (three seeds; percentages, mean $\pm$ sample SD). Frame F1/recall use official 20-ms labels. Event P/R/F1 match predicted and original 10-ms reference spans at IoU $\geq0.5$; short R is recall for reference events shorter than 100 ms. PS eval is retrospective.}
\label{tab:localization}
\setlength{\tabcolsep}{2.4pt}
\resizebox{\columnwidth}{!}{%
\begin{tabular}{lrrrrrrr}
\toprule
Arm & EER & Frame F1 & Frame R & Event P & Event R & Event F1 & Short R \\
\midrule
Hard frame & $11.123\pm0.020$ & $85.945\pm0.010$ & $82.420\pm0.341$ & $47.472\pm0.973$ & $66.763\pm0.931$ & $55.479\pm0.505$ & $60.957\pm1.788$ \\
Soft frame & $11.326\pm0.132$ & $85.657\pm0.135$ & $82.267\pm0.556$ & $47.242\pm0.777$ & $63.565\pm0.325$ & $54.199\pm0.572$ & $53.984\pm0.629$ \\
Phone unit & $11.237\pm0.224$ & $86.030\pm0.314$ & $83.904\pm0.817$ & $57.944\pm1.735$ & $43.823\pm0.697$ & $49.888\pm0.394$ & $15.694\pm1.375$ \\
Word unit & $12.514\pm0.182$ & $84.624\pm0.311$ & $82.657\pm0.594$ & $61.313\pm0.716$ & $39.665\pm0.459$ & $48.169\pm0.559$ & $11.730\pm0.221$ \\
\bottomrule
\end{tabular}}
\end{table}

\begin{table}[t]
\centering
\caption{Layer-6 PS-eval probes (three-seed mean $\pm$ SD, \%). Event metrics audit the same soft-frame and phone-unit checkpoints at saved PS-dev thresholds; context arms were not audited.}
\label{tab:anchors}
\scriptsize
\setlength{\tabcolsep}{2.2pt}
\begin{tabular}{lccc}
\toprule
Prediction / context & Frame EER & Event F1 & Short R \\
\midrule
Hard frame & $9.772\pm0.487$ & --- & --- \\
Soft frame & $9.591\pm0.284$ & $64.297\pm0.400$ & $68.722\pm1.510$ \\
Phone-unit output & $11.180\pm0.444$ & $52.291\pm0.296$ & $20.704\pm0.610$ \\
Frame + phone context & $9.656\pm0.410$ & --- & --- \\
Frame + duplicated frame & $9.928\pm0.541$ & --- & --- \\
Frame + fixed 80 ms & $10.176\pm0.661$ & --- & --- \\
\bottomrule
\end{tabular}
\end{table}

\subsection{Editing errors depend on the operating point}
Table~\ref{tab:interventions} reports the completed supplementary controls. At dev-EER thresholds, frame augmentation reduces GG FPR from 25.63\% to 5.48\%, while GF-core FNR rises from 8.98\% to 22.31\%. The paired changes are $-20.15$ percentage points (pp), interval $[-22.66,-17.81]$, and $+13.33$ pp, $[9.31,17.82]$. Phone models show the same direction. Initial dev controls also showed lower frame GG FPR (24.71\% to 4.39\%) and higher GF-core FNR (6.68\% to 12.52\%).

Calibration changes this pattern (Table~\ref{tab:operating}); GG excess FPR controls for unedited-genuine errors. At dev 5\% FPR, augmentation raises GF-core FNR by $13.31$ pp $[9.22,17.88]$ for frames and $16.48$ pp $[12.29,20.43]$ for phones. At dev 1\%, the changes are $1.11$ pp $[-2.40,4.76]$ and $1.85$ pp $[-3.09,6.33]$: neither establishes an increase or equivalence. Full source-FPR curves and G rates are released.

An intervention-specific Llama A follow-up froze 200 same-source-speaker cases (40 speakers), nine frame checkpoints, and PS-dev thresholds. At dev 5\% FPR, augmentation changes GG excess FPR by $-3.81$ pp (paired speaker/seed interval $[-5.09,-2.70]$) and GF-core FNR by $+4.62$ pp $[-3.15,11.87]$; Holm-adjusted bootstrap $p$ values are $<.001$ and $.236$, respectively. With 50-ms smoothing, original-model GF-core FNR instead changes by $-7.99$ pp $[-14.04,-2.10]$. These constructed cases are not official trials or pristine blind confirmation: broader Llama A transfer was inspected earlier, and texts/generators were unmatched.

Consistency gives no uniform improvement over augmentation. At dev-EER thresholds it raises frame GG FPR but lowers phone GG FPR, with the reverse direction for GF-core misses. The phone-minus-frame interaction in the change of GG excess FPR is $-0.96$ pp $[-2.93,0.75]$; it does not establish a combined advantage.

A post-hoc constructed GG/GF-core AUC check yields $0.9050\pm0.0065$, $0.9353\pm0.0118$, and $0.9336\pm0.0135$ for original, augmentation, and consistency frame models (three-seed mean $\pm$ SD). This is not official benchmark AUC or a selection metric: better ranking can coexist with more fixed-threshold misses.

\begin{table}[t]
\centering
\caption{Supplementary controls: 500 cases, 48 speakers, three-seed means (\%), original dev-EER thresholds. G/GG: unedited/genuine-splice FPR; GF/GF50: raw/50-ms-smoothed synthetic-core FNR. Final-layer models; lower is better.}
\label{tab:interventions}
\setlength{\tabcolsep}{4pt}
\begin{tabular}{llrrrr}
\toprule
Unit & Training & G & GG & GF & GF50 \\
\midrule
Frame & Original & 2.89 & 25.63 & 8.98 & 7.18 \\
Frame & Augment & 2.38 & 5.48 & 22.31 & 20.72 \\
Frame & +Consistency & 3.23 & 6.41 & 20.52 & 19.60 \\
Phone & Original & 2.37 & 27.80 & 13.12 & 14.76 \\
Phone & Augment & 1.97 & 4.33 & 29.95 & 29.61 \\
Phone & +Consistency & 1.99 & 3.47 & 31.66 & 31.83 \\
\bottomrule
\end{tabular}
\end{table}

\begin{table}[t]
\centering
\caption{The same controls at dev-calibrated 1\%/5\% FPR thresholds (three-seed means, \%). GG--G is genuine-splice FPR minus unedited genuine FPR; GF is synthetic-core FNR. Target-domain FPR is not fixed.}
\label{tab:operating}
\setlength{\tabcolsep}{4pt}
\begin{tabular}{llrrrr}
\toprule
& & \multicolumn{2}{c}{Dev 1\%} & \multicolumn{2}{c}{Dev 5\%} \\
Unit & Training & GG--G & GF & GG--G & GF \\
\midrule
Frame & Original & 8.48 & 33.90 & 21.61 & 9.90 \\
Frame & Augment & 1.06 & 35.01 & 2.86 & 23.21 \\
Frame & +Consistency & 1.20 & 34.95 & 2.95 & 21.48 \\
Phone & Original & 8.68 & 47.64 & 22.97 & 15.72 \\
Phone & Augment & 0.65 & 49.48 & 2.03 & 32.21 \\
Phone & +Consistency & 0.46 & 49.81 & 1.26 & 33.61 \\
\bottomrule
\end{tabular}
\end{table}

\subsection{Classification gains do not ensure better boundaries}
The layer-6 phone offset loss lowers mean PS eval EER (11.18\% to 10.69\%), but correcting the \emph{same} checkpoints lowers event F1 (50.47\% to 46.41\%; every seed) and worsens boundary error. The dev comparison also falls (68.03\% to 62.23\%). Classification improvement therefore does not establish localization improvement.

\subsection{Source selection does not determine deployment behavior}
Table~\ref{tab:transfer} shows a ranking reversal: layer 6 is best on PS development, whereas layer 12 is better on PS eval and both Llama archives; layer 6 reaches 56.582\% on HAD. We retain these source-locked results without selecting a layer from external labels.

At saved PS-development thresholds, final/layer-6/layer-12 FPR/FNR pairs are 63.92/13.80, 80.96/20.85, and 63.86/11.03\% on HAD; 28.26/56.00, 44.16/39.06, and 24.14/55.62\% on Llama A; and 39.07/38.10, 57.77/24.42, and 35.64/35.77\% on Llama B. The 36 seed records are released. Nonmatched reCFPRF gives 42.37/3.90\% on HAD, 37.22/46.51\% on A, and 56.56/24.42\% on B; HAD EER is 14.862\%.

\section{Scope and Limitations}
Findings concern tested encoders, alignments, heads, and edits. Alignment errors confound pooling; cross-corpus shifts mix language, synthesis, and editing. Non-speech and failed-alignment trials remain included. Three seeds cover training variability only; donor reuse, 400-ms inserts, and pointwise intervals limit inference. Grid-label events omit continuous-reference RangeEER; reCFPRF proposals and extraction timing remain unaudited.

\enlargethispage{2\baselineskip}
\section{Conclusion}
Tested phone and word units lose short-event recall despite some event-precision gains. Augmentation reduces benign-edit false alarms; synthetic-content misses depend on threshold and corpus, while consistency has no uniform benefit. Source-domain rankings do not determine cross-corpus behavior. Paired GG/GF controls, localization, and source-locked transfer are needed to assess these designs; conclusions do not extend to all anchors or consistency methods.

\noindent\textbf{AI disclosure.} ChatGPT assisted wording; the authors verified all claims and results.

\clearpage
\begingroup
\footnotesize
\balance

\endgroup
\end{document}